\documentclass[aps,prl,reprint,nofootinbib,preprintnumbers,superscriptaddress]{revtex4-2}
\pdfoutput=1

\usepackage[utf8]{inputenc}
\usepackage[T1]{fontenc}
\usepackage{graphicx}
\usepackage{amsmath}
\usepackage{hyperref}

\newcommand{\diag}{\operatorname{diag}}

\newcommand{\Dmq}{\Delta m^2}
\newcommand{\Eps}{\varepsilon}
\newcommand{\Epx}{\mathcal{E}}

\begin{document}

\title{On the effect of NSI in the present determination of the mass ordering}

\author{Ivan Esteban,}
\email{ivan.esteban@fqa.ub.edu}
\affiliation{Departament de Fis\'{\i}ca Qu\`antica i
  Astrof\'{\i}sica and Institut de Ciencies del Cosmos, Universitat de
  Barcelona, Diagonal 647, E-08028 Barcelona, Spain}

\author{M.~C.~Gonzalez-Garcia,}
\email{maria.gonzalez-garcia@stonybrook.edu}
\affiliation{Departament de Fis\'{\i}ca Qu\`antica i Astrof\'{\i}sica
  and Institut de Ciencies del Cosmos, Universitat de Barcelona,
  Diagonal 647, E-08028 Barcelona, Spain}
\affiliation{C.N.~Yang Institute for Theoretical Physics, State
  University of New York at Stony Brook, Stony Brook, NY 11794-3840,
  USA}
\affiliation{Instituci\'o Catalana de Recerca i Estudis Avan\c{c}ats
  (ICREA), Pg.\ Lluis Companys 23, 08010 Barcelona, Spain}

\author{Michele Maltoni}
\email{michele.maltoni@csic.es}
\affiliation{Instituto de F\'{\i}sica Te\'orica UAM/CSIC, Calle de
  Nicol\'as Cabrera 13--15, Universidad Aut\'onoma de Madrid,
  Cantoblanco, E-28049 Madrid, Spain}

\begin{abstract}
  In a recent work by Capozzi~\textit{et al}~\cite{Capozzi:2019iqn},
  it is observed that the introduction of non-standard neutrino-matter
  interactions considerably relaxes the preference of T2K and NO$\nu$A
  for normal over inverted mass ordering observed in the standard
  three-neutrino scenario. Motivated by this, in this note we update
  our previous global fit to investigate whether such result still
  holds once the information of solar, atmospheric and reactor
  experiments is taken into account. We find that the non-standard
  parameters responsible for the improvement of the inverted ordering
  fit to T2K and NO$\nu$A data are \emph{not} compatible with the
  other oscillation experiments, and that the preference for NO is
  restored.
\end{abstract}


\preprint{IFT-UAM/CSIC-20-53, YITP-SB-2020-5}

%\arxivnumber{XXXX.XXXXX}

\maketitle

\section{Introduction}

The unambiguous determination of the neutrino mass ordering is one of
the primary goals of forthcoming long-baseline (LBL) neutrino
experiments. The technical requirements needed to achieve it have been
widely studied in the context of the standard three-neutrino
oscillation framework, so that success is guaranteed as long as no
``unexpected'' physical phenomenon takes place.  On the other hand, in
the presence of New Physics the capability of a given experiment to
resolve the neutrino mass ordering can be severely reduced. For
example, in the context of non-standard neutrino-matter interactions
(NSI) a new parameter degeneracy is present, which involves a change
in the octant of the solar mixing angle (thus leading to the
appearance of a new region characterized by $\theta_{12}$ in the
second octant, the so-called LMA-Dark (LMA-D)
solution~\cite{Miranda:2004nb}) as well as a change in the neutrino
mass ordering (\textit{i.e.}, sign of
$\Dmq_{31}$)~\cite{Gonzalez-Garcia:2013usa, Bakhti:2014pva,
  Coloma:2016gei}.  This ``generalized mass ordering
degeneracy''~\cite{Coloma:2016gei} cannot be resolved by Earth-based
oscillation experiments alone, and therefore it undermines the
capability of \emph{any} LBL experiment to establish the neutrino mass
ordering. The degeneracy is only approximate once experiments
observing neutrinos which have traveled through matter with variable
chemical composition, such as the Sun, are included into the fit.
Still the LMA-D region, and hence the corresponding inversion in the
mass ordering, remains a valid solution in the global analysis of
oscillation data for a broad spectrum of NSI with
quarks~\cite{Esteban:2018ppq}. However, the appearance of the
LMA-D/flip-ordering solution requires pretty large values of the NSI
parameters, which lead to sizable effects on non-oscillation neutrino
experiments such as COHERENT~\cite{Akimov:2017ade}. This means that
the generalized mass-ordering degeneracy can be resolved (at least in
principle) by combining data from both oscillation and scattering
neutrino experiments~\cite{Coloma:2019mbs}.

Even if the LMA-D solution is ruled out, the introduction of NSI may
still pose a threat to the sensitivity of LBL experiments by simply
allowing for small but potentially dangerous deformations of the
standard three-neutrino oscillations. In this case, the NSI parameters
need not be particularly large, hence the possibility of disentangling
them from genuine neutrino masses and mixing is not guaranteed. A
detailed analysis of such situation using all the solar, atmospheric,
reactor and accelerator neutrino data available at the end of 2018 was
presented in Ref.~\cite{Esteban:2019lfo}. In that work it was found
that \emph{no} further parameter degeneracy beyond the LMA-D
generalized mass-ordering one is induced by NSI, and that the $\sim
2\sigma$ preference for normal over inverted ordering observed in the
standard three-neutrino scenario~\cite{Esteban:2018azc} is not
affected by NSI as long as the LMA-D solution is neglected.

A similar analysis, updated with the data released in summer 2019 but
limited to the T2K and NO$\nu$A accelerator experiments, was later
presented in Ref.~\cite{Capozzi:2019iqn}, leading however to different
conclusions. The authors found that the current sensitivity to the
mass ordering~\cite{nufit-4.1} is completely washed out once NSI are
introduced in the fit, even without considering the LMA-D region.  The
authors ascribe the discrepancy between their result and those in
Ref.~\cite{Esteban:2019lfo} to the different data sets used in the two
analyses, in particular for what concerns the T2K and NO$\nu$A
results. Specifically, the slightly newer data used in
Ref.~\cite{Capozzi:2019iqn} are found to significantly favor a
non-zero value of $|\Eps_{e\tau}|$ when inverted ordering is
considered, hence improving the quality of the corresponding fit and
removing the tension with respect to normal ordering.

In view of this recent development, in this note we update our former
analysis~\cite{Esteban:2019lfo} to account for the newer T2K and
NO$\nu$A data included in Ref.~\cite{Capozzi:2019iqn}. We perform
first an analysis including only $\Eps_{e\tau}$ so to directly address
the impact of this parameter on the ordering determination once the
information from all experiments is taken into account. Second we
update our global analyses including all NSI couplings with the new
LBL data samples as well as the timing and energy information from the
COHERENT experiment as detailed in Ref.~\cite{Coloma:2019mbs}.  In
Sec.~\ref{sec:discussion} we describe the results of our analysis and
in Sec.~\ref{sec:conclusions} we draw our conclusions.

\section{Discussion}
\label{sec:discussion}

We are going to consider NSI affecting neutral-current processes
relevant to neutrino propagation in matter. The coefficients
accompanying the relevant operators are usually parametrized in the
form:
\begin{equation}
  \label{eq:NSILagrangian}
  \begin{aligned}
    \mathcal{L}_\text{NSI}
    &= -2\sqrt2 G_F
    \sum_{f,\alpha,\beta} \Eps_{\alpha\beta}^f
    (\bar\nu_\alpha\gamma^\mu P_L\nu_\beta)
    (\bar f\gamma_\mu  f) \,,
  \end{aligned}
\end{equation}
where $G_F$ is the Fermi constant, $\alpha$ and $\beta$ are flavor
indices, and $f$ is a SM charged fermion. In this notation,
$\Eps_{\alpha\beta}^f$ parametrizes the strength of the vector part of
the new interactions (which are the ones entering the neutrino matter
potential) with respect to the Fermi constant, $\Eps_{\alpha\beta}^f
\sim \mathcal{O}(G_X/G_F)$.
In this framework, the evolution of the neutrino and antineutrino
flavor state during propagation is governed by the Hamiltonian:
\begin{equation}
  H^\nu = H_\text{vac} + H_\text{mat}
  \quad\text{and}\quad
  H^{\bar\nu} = ( H_\text{vac} - H_\text{mat} )^* \,,
\end{equation}
where $H_\text{vac}$ is the vacuum part which in the flavor basis
$(\nu_e, \nu_\mu, \nu_\tau)$ reads
\begin{equation}
  H_\text{vac} = \frac{1}{2E_\nu} U_\text{vac} \cdot
  \diag(0, \Dmq_{21}, \Dmq_{31}) \cdot U_\text{vac}^\dagger \,.
\end{equation}
Here $U_\text{vac}$ denotes the three-lepton mixing matrix in
vacuum~\cite{Pontecorvo:1967fh, Maki:1962mu, Kobayashi:1973fv} which
we parametrize following the conventions of
Ref.~\cite{Coloma:2016gei}.

If all possible operators in Eq.~\eqref{eq:NSILagrangian} are added to
the SM Lagrangian, the matter part $H_\text{mat}$ is a function of the
number densities $N_f(x)$ of the fermions $f$ in the matter along the
trajectory:
\begin{equation}
  \label{eq:Hmat}
  H_\text{mat} = \sqrt{2} G_F N_e(x)
  \begin{pmatrix}
    1 + \Epx_{ee}(x) & \Epx_{e\mu}(x) & \Epx_{e\tau}(x) \\
    \Epx_{e\mu}^*(x) & \Epx_{\mu\mu}(x) & \Epx_{\mu\tau}(x) \\
    \Epx_{e\tau}^*(x) & \Epx_{\mu\tau}^*(x) & \Epx_{\tau\tau}(x)
  \end{pmatrix}
\end{equation}
where the ``$+1$'' term in the $ee$ entry accounts for the standard
contribution, and
\begin{equation}
  \label{eq:epx-nsi}
  \begin{split}
    \Epx_{\alpha\beta}(x)
    &= \sum_{f=u,d} \frac{N_f(x)}{N_e(x)} \Eps_{\alpha\beta}^f
    \\
    &= \Eps_{\alpha\beta}^e + 2\Eps_{\alpha\beta}^u + \Eps_{\alpha\beta}^d
    + Y_n(x) \big(2\Eps_{\alpha\beta}^d + \Eps_{\alpha\beta}^u \big)
  \end{split}
\end{equation}
describes the non-standard part. $Y_n(x) \equiv N_n(x) \big/ N_e(x)$
is the composition-dependent neutron abundance and we have used that
matter neutrality implies $N_e(x) = N_p(x)$.  For experiments where
neutrinos travel in the Earth matter, like LBL experiments, one can
safely set $Y_n(x)$ to a fixed value $Y_n^\oplus$ which is not very
different from one.\footnote{The PREM model~\cite{Dziewonski:1981xy}
  fixes $Y_n=1.012$ in the mantle and $Y_n = 1.137$ in the core, so
  that for atmospheric and LBL neutrino experiments one can set it to
  an average value $Y_n^\oplus = 1.051$ all over the Earth.}  Within
this approximation, the analysis of atmospheric and LBL neutrinos
holds for any combination of NSI with up and down quarks as well as
electrons, and it can be performed in terms of the effective NSI
couplings $\Eps_{\alpha\beta}^\oplus$, given by Eq.~\eqref{eq:epx-nsi}
with a constant $Y_n^\oplus$, which play the role of phenomenological
parameters.

\begin{figure*}\centering
  \includegraphics[width=\linewidth]{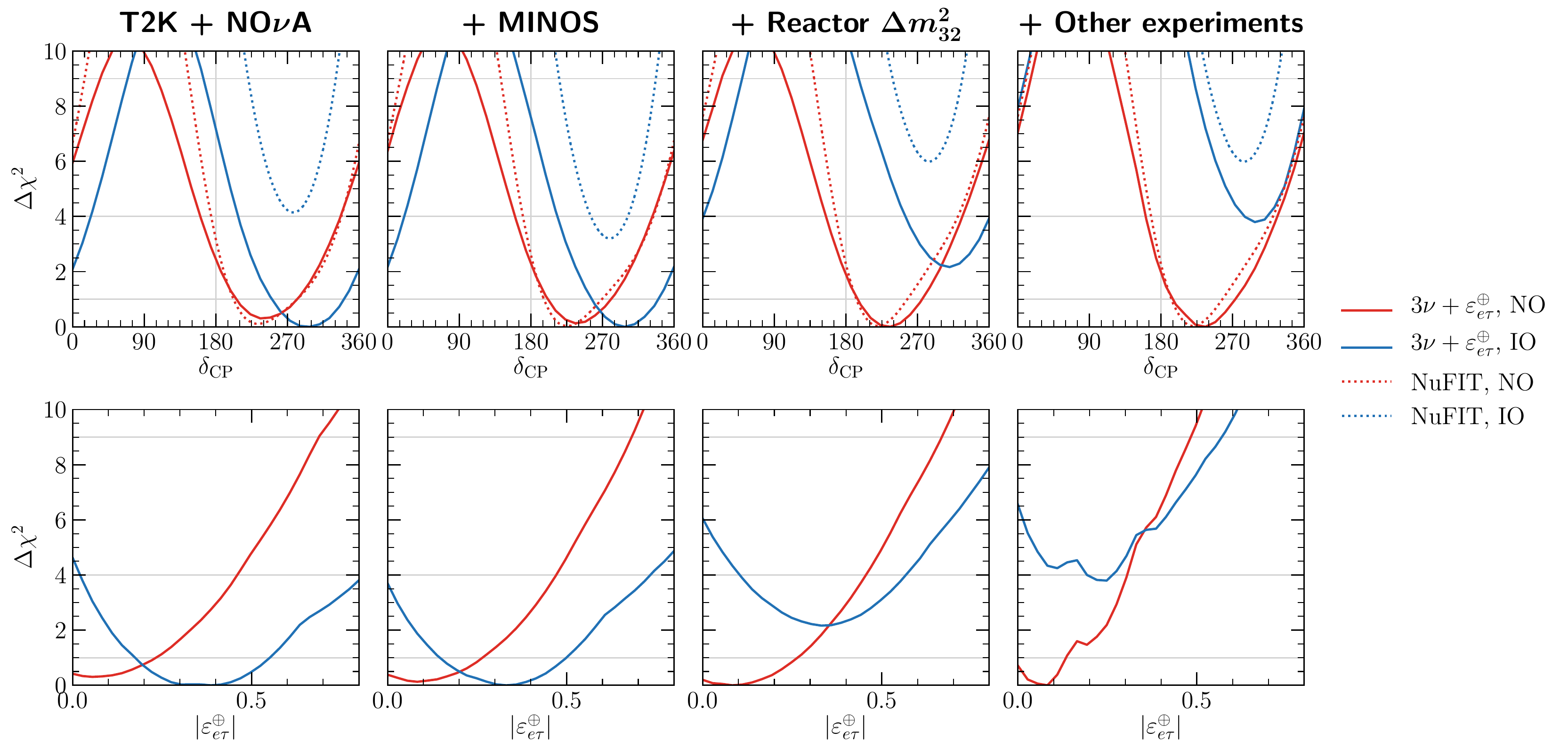}
  \caption{$\Delta\chi^2$ as a function of $\delta\text{CP}$ (above)
    and $|\Eps_{e\tau}^\oplus|$ (below) obtained from the combined
    analysis of different set of oscillation data (as labeled on the
    top of each panel) after marginalizing over the undisplayed
    parameters.  In each panel we plot the curves obtained
    marginalizing separately in NO (red) and IO (blue). For the sake
    of comparison we also plot the corresponding $\Delta\chi^2$ from
    the $3\nu$ oscillation analysis with the SM matter potential
    (labeled ``NuFIT'' in the figure).}
  \label{fig:epset}
\end{figure*}

In order to better address the sensitivity loss to the ordering found
in Ref.~\cite{Capozzi:2019iqn}, we start by performing our analysis in
the same scenario studied there: oscillations of three massive
neutrinos in the presence of a single NSI effective complex parameter
$\Eps_{e\tau}^\oplus$.
The results are presented in Fig.~\ref{fig:epset}, where we plot the
$\Delta\chi^2$ from the combined analysis of different sets of
oscillation data as a function of $\delta_\text{CP}$ (above) and
$|\Eps_{e\tau}^\oplus|$ (below) after marginalizing over the
undisplayed parameters.  In each panel we plot the curves obtained
marginalizing separately in NO (red) and IO (blue). For the sake of
comparison in the upper panels we also plot the corresponding
$\Delta\chi^2(\delta_\text{CP})$ from the $3\nu$ oscillation analysis
with the SM matter potential (labeled ``NuFIT'' in the figure).

In the left panels of Fig.~\ref{fig:epset} we present the results of
the analysis including only the updated results of T2K and NO$\nu$A.
In this analysis $\Dmq_{12}$ and $\theta_{12}$ are fixed to their best
fit value determined by solar and KamLAND data, and a bias for
$\theta_{13}$ is included to account for the constraints from the
medium-baseline (MBL) reactor experiments
Double-CHOOZ~\cite{dc:cabrera2016}, Daya-Bay~\cite{An:2016ses} and
RENO~\cite{reno:eps2017}.  The curves shown here can be directly
compared with the corresponding ones in the left panels of Fig.~3 in
Ref.~\cite{Capozzi:2019iqn}, with which we find a good agreement. In
particular we note the $\sim 2.1\sigma$ preference for NO from the
analysis of T2K and NO$\nu$A without NSI, clearly visible from the
dotted curves, is completely washed out once $\Eps_{e\tau}^\oplus$ NSI
are introduced, as shown by the solid lines.  Also from the lower-left
panel we find that the best fit in IO corresponds to a non-zero value
of $\Eps_{e\tau}^\oplus\sim 0.4$ (well in agreement with
Ref.~\cite{Capozzi:2019iqn}) which is preferred over the SM null value
by more than $2\sigma$.

The other panels in Fig.~\ref{fig:epset} quantify how this result is
affected by the inclusion of additional data samples. In the
center-left panels the information from the MINOS LBL experiment is
included. This leads to a slight improvement of IO with respect to NO
in the standard $3\nu$ scenario, but the main conclusion about the
total wash-out of the preference for NO in the presence of
$\Eps_{e\tau}$ still holds and the corresponding projections over
$|\Eps_{e\tau}^\oplus|$ are barely changed.

In the center-right panels we also add the information on $\Dmq_{31}$
from the MBL reactor experiments~\cite{dc:cabrera2016, An:2016ses,
  reno:eps2017}, so that we no longer assume a bias on $\theta_{13}$
but instead we combine the full data from LBL accelerator and MBL
reactor experiments. First of all, we note that the inclusion of the
$\Dmq_{31}$ information from MBL reactors adds to the preference for
NO in the standard $3\nu$ oscillation scenario. This is due to the
well known fact that the precise determination of the oscillation
frequencies in $\nu_\mu$ disappearance at LBL experiments and $\nu_e$
disappearance in reactor experiments yields information on the sign of
$\Dmq_{31}$. With the present MBL data it adds substantially to the
preference for NO which in the $3\nu$ oscillation scenario is favored
at the $\sim 2.5\sigma$ level. But more importantly, given the short
baselines of these reactor experiments which renders them practically
insensitive to matter effects, this extra preference for NO cannot be
washed out by the inclusion of NSI.  Hence from the center-right
panels we see that NO is favored over IO at the $\sim 1.4\sigma$ level
even in the presence of $\Eps_{e\tau}$. Consistently the shape of the
dependence of the $\chi^2$ on $|\Eps_{e\tau}^\oplus|$ for IO does not
change by the inclusion of the MBL data, although its minimum value is
shifted.

\begin{figure*}\centering
  \includegraphics[width=\linewidth]{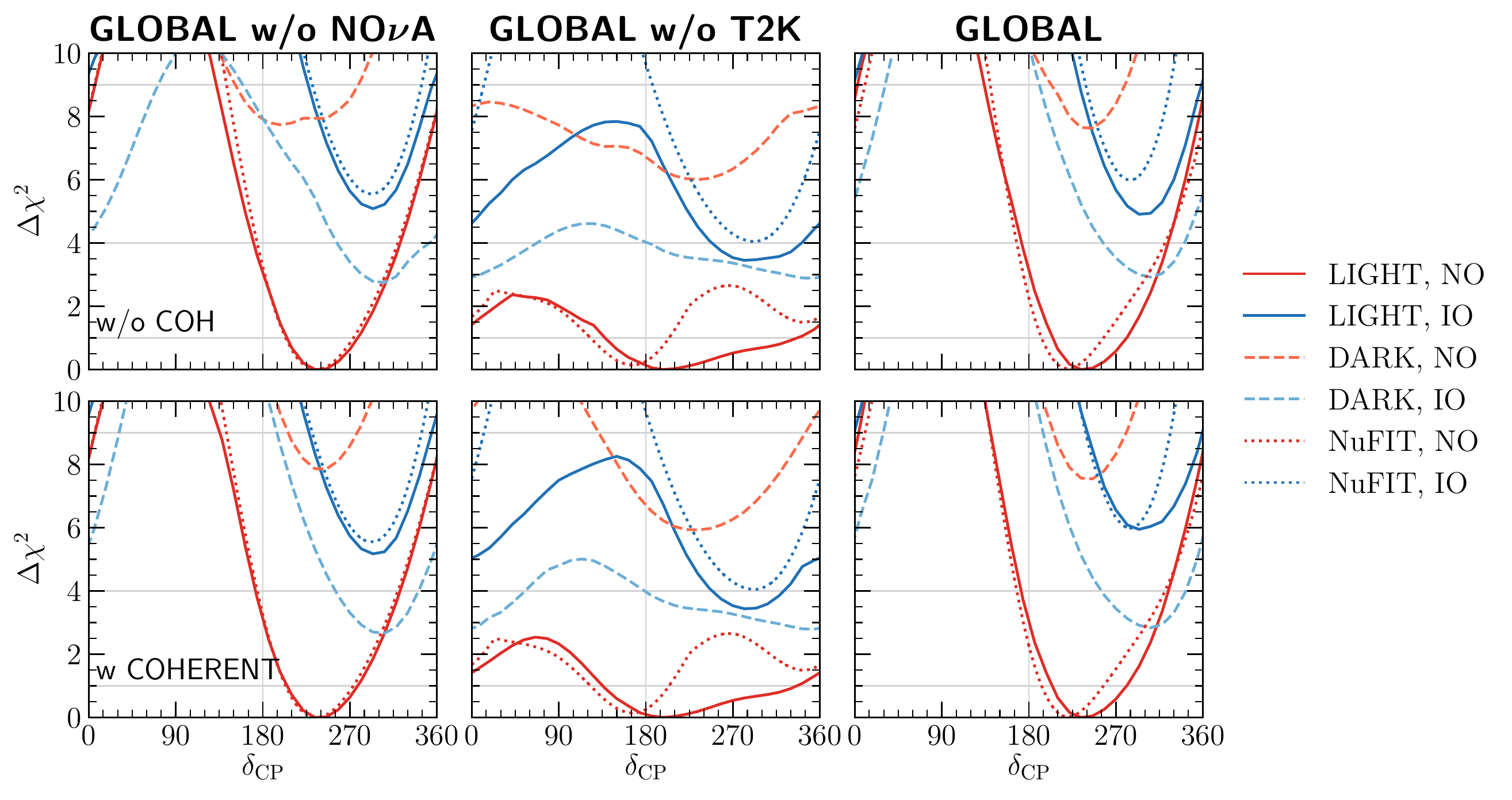}
  \caption{$\Delta\chi^2_\text{GLOB}$ as a function of
    $\delta_\text{CP}$ after marginalizing over all the undisplayed
    parameters, for different combination of experiments.  In all
    panels we include $\text{SOLAR} + \text{KamLAND} + \text{MBL-REA}
    + \text{MINOS}$ to which we add T2K (left), NO$\nu$A (center) and
    $\text{T2K} + \text{NO$\nu$A}$ (right).  The corresponding lower
    panels also include the updated constraints from COHERENT.  The
    different curves are obtained by marginalizing within different
    regions of the parameter space, as detailed in the legend.}
  \label{fig:fullNSI}
\end{figure*}

Finally, in the rightmost panels of Fig.~\ref{fig:epset} we add the
information from all other experiments and perform a global analysis
where the the LBL and MBL results are explicitly combined with the
data from solar neutrino experiments, IceCube and its sub-detector
DeepCore, and Super-Kamiokande (SK) atmospheric data (see
Ref.~\cite{Esteban:2019lfo} for the details on the analysis and the
references to the data included).\footnote{When including the
  constraints from non-terrestrial experiments the analysis is
  performed in terms of the Lagrangian parameters
  $\Eps_{\alpha\beta}^f$ introduced in Eq.~\eqref{eq:NSILagrangian},
  and then projected onto the effective $\Eps_{\alpha\beta}^\oplus$
  parameters as described in Ref.~\cite{Esteban:2019lfo}.
  Quantitatively the effect is not very different from adding a prior
  on the modulus of the $\Eps_{\alpha\beta}^\oplus$ parameters
  according to the projections shown in Fig.~9 of
  Ref.~\cite{Esteban:2018ppq}.  Tables of such projections for
  specific combinations can be provided upon request to the authors.}
As seen in the figure, once the results from all experiments are
combined, the preference of NO at the $2\sigma$ level is recovered
even in the presence of the $\Eps_{e\tau}$ NSI. From the lower-right
panel we see that the analysis in IO still favors a non-zero (though
smaller) $|\Eps_{e\tau}^\oplus| \sim 0.25$ though its significance is
reduced to $\Delta\chi^2 \sim 2.5$.  Indeed the curve shows a second
almost degenerate minima at smaller $|\Eps_{e\tau}^\oplus|\sim 0.1$ as
a consequence of the tension of the larger value favored by
T2K+NO$\nu$A and the zero value favored by the rest of the
experiments.

In summary, we conclude that the non-standard parameters responsible
for the improvement of the inverted ordering fit to T2K and NO$\nu$A
data, as observed in Ref.~\cite{Capozzi:2019iqn} and reproduced in our
own analysis shown in the left panel of Fig.~\ref{fig:epset}, are
\emph{not} compatible with the other oscillation experiments, and that
the preference for NO is restored once the data from those experiments
is included in the analysis.

We next move to study the effect of the updated LBL samples on the
conclusions of our analysis in Ref.~\cite{Esteban:2019lfo} about the
LMA-D degeneracy. To this aim we perform an analysis including all the
NSI parameters, not just $\Eps_{e\tau}$.  The results are shown in
Fig.~\ref{fig:fullNSI} which can be directly compared with Fig.~3 of
Ref.~\cite{Esteban:2019lfo}.
In this figure we plot the one-dimensional $\chi^2(\delta_\text{CP})$
function obtained from $\chi^2_\text{GLOB}$ after marginalizing over
the ten undisplayed parameters. In the left, central and right panels
we focus on the GLOBAL analysis including T2K, NO$\nu$A, and
T2K+NO$\nu$A respectively --~so the data samples included in the
analysis corresponding to the right panels of Figs.~\ref{fig:epset}
and~\ref{fig:fullNSI} are the same.
In each panel we plot the curves obtained marginalizing separately in
NO (red) and IO (blue) and within the $\text{LIGHT}$ (full lines) and
$\text{DARK}$ (dashed lines) solutions. For the sake of comparison we
also plot the corresponding $\chi^2(\delta_\text{CP})$ from the $3\nu$
oscillation analysis with the SM matter potential (dotted lines
labeled ``NuFIT''). In the lower panels we also include the timing and
energy information from the COHERENT experiment as detailed in
Ref.~\cite{Coloma:2019mbs}.
Compared to Fig.~3 of Ref.~\cite{Esteban:2019lfo} we find that, even
though there are minor quantitative differences (especially in the
central panel), the main conclusions with respect to the status of the
LMA-D ordering degeneracy remain unchanged. In particular in the
global analysis (right panel of Fig.~\ref{fig:fullNSI}) the DARK-IO
solution is still allowed below $2\sigma$, but it has become slightly
more disfavored and it is now at $\Delta\chi^2\sim 3$ (in
Ref.~\cite{Esteban:2019lfo} it was $\Delta\chi^2\sim 2$) with respect
to the best fit LIGHT-NO.

We finish by noticing that comparing the results for the LIGHT
solutions in the global analysis in the upper-right panel of
Figs.~\ref{fig:fullNSI} with those of the corresponding $3\nu +
\Eps_{e\tau}$ in the upper-right panel of Fig.~\ref{fig:epset}, we see
that enlarging the number of NSI couplings included in the analysis
increases the preference for NO. This may seem counterintuitive.  The
reason is that allowing non-vanishing $\Eps_{e\mu}$ results in an
improvement of the fit in both NO and IO with respect to the standard
3$\nu$ fit, a result also pointed out in Ref.~\cite{Capozzi:2019iqn}.
In the global analysis, the improvement is slightly more significant
for NO, hence the increase of the preference for NO in this case. We
also see that the minima of the blue full and dotted lines in the
lower right panel lay at the same $\Delta\chi^2$. In other words, once
the updated constraints from COHERENT are included we find the same
preference for NO for LIGHT solutions than in the standard $3\nu$
oscillation scenario.

\section{Conclusions}
\label{sec:conclusions}

In this note we have updated our former
analysis~\cite{Esteban:2019lfo} by including the newer T2K and
NO$\nu$A data.  We have found that:
\begin{itemize}
\item when only the LBL accelerator data from T2K and NO$\nu$A are
  considered, the introduction of a non vanishing NSI $\Eps_{e\tau}$
  considerably relaxes the preference for normal over inverted
  ordering found in the standard three-neutrino oscillation scenario,
  in line with what was pointed out in Ref.~\cite{Capozzi:2019iqn};

\item on the other hand, once the information from other oscillation
  experiments (solar, atmospheric, reactor as well as MINOS
  experiments) is taken into account the large $|\Eps_{e\tau}|$ values
  responsible for the improvement of the fit in the IO become
  disfavored, thus restoring the preference for NO observed in the
  standard scenario.

\item the status of the LMA-D degeneracy is only mildly affected by
  the inclusion of the updated LBL data and COHERENT timing and energy
  information.
\end{itemize}
In summary, the updated results of the global analysis reconfirm the
conclusions of Ref.~\cite{Esteban:2019lfo}.

\section*{Acknowledgments}

This work was supported by the MINECO grant FPA2016-76005-C2-1-P, by
the MINECO FEDER/UE grants FPA2016-78645-P, by USA-NSF grants
PHY-1620628, by EU Network FP10 ITN ELUSIVES
(H2020-MSCA-ITN-2015-674896), by the ``Severo Ochoa'' program grant
SEV-2016-0597 of IFT and by AGAUR (Generalitat de Catalunya) grant
2017-SGR-929.  IE acknowledges support from the FPU program fellowship
FPU15/0369.

\bibliography{references}

\end{document}